# The Price of Anarchy (POA) of network coding and routing based on average pricing mechanism

Gang Wang
School of Electronics and Information Engineering
Beihang University
Beijing, China
gwang@buaa.edu.cn

Xia Dai
School of Electronics and Information Engineering
Beihang University
Beijing, China
sunnydora01@hotmail.com

*Abstract*—The congestion pricing is an efficient allocation approach to mediate demand and supply of network resources. Different from the previous pricing using Affine Marginal Cost (AMC), we focus on studying the game between network coding and routing flows sharing a single link when users are price anticipating based on an Average Cost Sharing (ACS) pricing mechanism. We characterize the worst-case efficiency bounds of the game compared with the optimal, i.e., the price-of anarchy (POA), which can be low bound 50% with routing only. When both network coding and routing are applied, the POA can be as low as 4/9. Therefore, network coding cannot improve the POA significantly under the ACS. Moreover, for more efficient use of limited resources, it indicates the sharing users have a higher tendency to choose network coding.

I. INTRODUCTION

The current Internet is frequent that a group of different users share the same link transforming information in a wide variety of real network situations. The congestion pricing, as an effective approach in terms of improving the efficiency of network resource allocation, is set to mediate demand and supply of network resources.

Since the seminal paper by Ahlswede et al. [1], network coding has shown great potential for improving network throughput in communication networks. However, game theoretic analysis on the performance of network coding has received more and more attention recently, e.g., in [2]-[7]. All results in [2]-[4] focus on the case of intra-session network coding performed by jointly encoding multiple packets from the same user, whereas inter-session network coding encoding packets from different users is considered in [5-7]. Moreover, a game theoretic analysis for inter-session network coding of combining network coding and routing flows on a single bottleneck link is investigated in [5-7]. In addition, a key performance metric named the price of anarchy (POA) for analyzing the system, is the ratio between the total payoffs at the worst Nash equilibrium of this game and the efficient payoffs, i.e., the maximum feasible total payoffs. The total payoffs are denoted by aggregate surplus, i.e., aggregate utility less aggregate cost. Obviously, it indicates that a higher POA denotes a smaller efficiency loss. All the results in [7] are based on the Affine Marginal Cost (AMC) pricing mechanism similar to [8].

An Average Cost Sharing (ACS) pricing mechanism noted by Vasilis Ntranos in [9] is more robust that users cannot benefit from merging or splitting their rates than the AMC mechanism generally, which is more reasonable and fairness to each sharing user. More specifically, both the number of sharing users and the total rates through the sharing link have been taken into consideration.

In this paper, we focus on studying the game between network coding and routing flows sharing a single link when users are price anticipating based on the ACS pricing mechanism. The innovative design of ACS pricing function is affected by the sum of all users' decisions and the total number of users sharing the same technique (network coding or routing). Each user's cost share charged by network is proportional to its action (transmission rate). The network aggregate surplus is formulated by the total utility of all users minus the total network cost of them. The key contributions of this paper is as follows: 1) A complete analysis of a sharing single link system with both network coding and routing flows is established based on ACS pricing mechanism. 2) POA is calculated under the two pricing circumstances, i.e., non-discriminatory pricing with routing-only users and discriminatory pricing with both network coding and routing.

The following is the outline of the remainder of this paper. In Section II, we achieve the results of the average cost sharing pricing mechanism for a single link on resource allocation games with routing-only. Then the case when some users can jointly perform inter-session network coding will be considered with extended results in Section III. Finally, conclusions are discussed in Section IV.

II. ACS PRICING MECHANISM ----ROUTING GAME

*A. Problem formulation*

In this section, we consider a framework with a single shared link $(w, z)$ to allocate network capacity efficiently among a collection of $n$ routing-only users, i.e., all data that node $w$ receive are simply forwarded to node $z$, as shown in Fig. 1. The set of users is denoted by $\mathcal{N} = \{1,2,...,n\}$. Each user $i \in \mathcal{N}$ is endowed with a utility function $\mathcal{U}_i(x_i)$ depending on their desired data transmission rate $x_i$ from its sender $u_i$ to its receiver $v_i$. In addition, let $q = \sum_{i=1}^{n} x_i$ denote the total rate allocated at the link, and the link is overloaded when the sum of rates $q$ through it exceeds the fixed capacity $C$ of the link. A cost share for the average queue experienced by user $i$ is denoted by $\mathcal{P}_i(x_i)$. Therefore, $\mathcal{P}_i(x_i)$ is the reduction in user $i$'s utility due to the network congestion. Generally, $\mathcal{U}_i(x_i)$ and $\mathcal{P}_i(x_i)$ are measured in the same monetary units. Moreover, $\mathcal{U}_i(x_i)$ is the monetary value to each user $i$ of a rate allocation $x_i$, while $\mathcal{P}_i(x_i)$ is a monetary cost for congestion at the link to each user $i$ of a rate allocation $x_i$ charged by network, respectively.

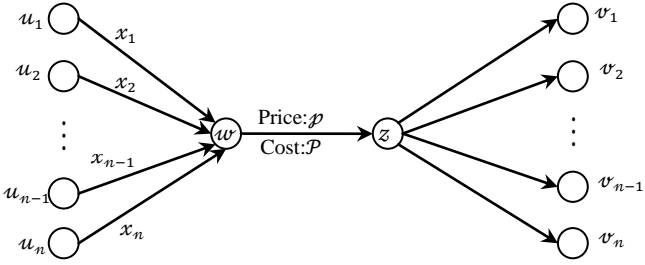

Figure 1. $n$ routing-only users sharing a single link

Next, we make the following assumptions regarding $\mathcal{U}_i(x_i)$ and $\mathcal{P}_i(x_i)$.

*Assumption 1*: There is a differentiable, convex, and non-decreasing price function $p(z)$ over $z \geq 0$, with $p(z) \geq 0$ and $p(z) \to \infty$ as $z \to \infty$, and we suppose the price function is linear for simplicity.

*Assumption 2*: For each transmission rate $x_i \geq 0, i \in \mathcal{N}$, the cost function for the $i^{th}$ user is modeled as $\mathcal{P}_i(x_i) = x_i \cdot p_{x_i}(z)$ subjected to $z \geq 0$. Here $\mathcal{P}_i(x_i)$ is convex and non-decreasing. Moreover, let $\sum_i \mathcal{P}_i(x_i)$ denote the cost incurred at the link when the total allocated rate is $q \geq 0$. Similarly, $\sum_i \mathcal{P}_i(x_i)$ is also convex and non-decreasing.

*Assumption 3*: For each $i \in \mathcal{N}$, utility function $\mathcal{U}_i(x_i)$ is concave, non-negative, strictly increasing, and differentiable. Moreover, for simplicity and generality, suppose the utility function is also linear for all users. That is $\mathcal{U}_i(x_i) = r_i x_i$, $\forall i \in \mathcal{N}$, where utility parameter $r_i > 0$ for all users.

In general, since the utility functions are local to the users and are not known at each link, the efficient resource allocation needs to be done via pricing. Therefore, the pricing mechanism applied is very significant for the system.

*Proposition 1*: Suppose assumption 1 holds. There exist two techniques for data transmission on the shared single link: network coding and routing. We define the following ACS pricing mechanism referring to [9]:

$$p\left(\frac{\sum_j x_j}{n_\ell}\right), \quad (1)$$

where $n_\ell$ denotes the number of users sharing one of the two independent single links (routing or network coding), and $\sum_j x_j$ is the sum of rates for each user $j$ through the single link with the same technique, $j = 1,2,\ldots,n_\ell$.

Given the rate vector $\boldsymbol{x} = (x_1, \ldots, x_n)$ from the users, the capacity of single communication link $(w, z)$ is supposed as $C = 1$, then $q = \sum_{i=1}^{n} x_i \leq C = 1$. For simplicity and generality, we assume that $1 \geq x_1 \geq x_2 \geq \ldots x_{n-1} \geq x_n \geq 0$ in the rest of this paper.

In this scenario, we only take the routing into account. Thus we set a single scalable price for the shared link to all the routing users:

$$\mu(x) = p\left(\frac{\sum_{i=1}^{n} x_i}{n}\right), \quad (2)$$

where $q = \sum_{i=1}^{n} x_i \leq 1$ is the sum of rates from all the routing users, $n$ is the number of routing users sharing the same link $(w, z)$. Obviously, the price is affected by the sum of all routing users' rates and the total number of routing users sharing the same link. Furthermore, the cost of each user $i \in \mathcal{N}$ for its transmission rate $x_i$ is $\mathcal{P}_i(x_i) = x_i \cdot \mu(x)$.

Efficiency is defined in terms of the aggregate value of a network allocation in [8]. Similarly, we also analyze the performance of the pricing mechanism from the point of efficient allocation, which can be characterized one via maximizing aggregate surplus as an optimal solution:

$$\text{maximize } \sum_i \mathcal{U}_i(x_i) - \sum_i \mathcal{P}_i(x_i), \quad (3)$$

$$\text{subject to } x_i \geq 0, \ i = 1, \ldots, n. \quad (4)$$

The above objective function (3) is referred as the aggregate surplus.

Then we should consider the surplus of each user $i \in \mathcal{N}$ for a price $\mu(x) \geq 0$, where they will play a game to acquire a share of link because users are anticipating the effect of their rates on the resulting price. The notation $\boldsymbol{x}_{-i}$ is denoted the vector of all rates chosen by users other than $i$. Then given $\boldsymbol{x}_{-i}$, each user $i$ will choose $x_i \geq 0$ to maximize its surplus:

$$\max_{x_i \geq 0} \mathcal{S}(x_i, \boldsymbol{x}_{-i}) = \max_{x_i \geq 0} [\mathcal{U}_i(x_i) - \mathcal{P}_i(x_i)], \ i = 1, \ldots, n. \quad (5)$$

In fact, the decision made by user $i$ depends on the rates selected by other users, leading to a resource allocation game among all routing users. A normal form of the game $G_1$ is given as follows:

$$G_1 = \langle \mathcal{N}, \{x_i\}, \{\mathcal{S}_i(x_i, \boldsymbol{x}_{-i})\} \rangle, \quad (6)$$

where $\mathcal{N} = \{1,2,\ldots,n\}$ is the set of players (sharing users), the rate $x_i$ is the strategy of each player $i \in \mathcal{N}$, and $\{\mathcal{S}_i(x_i, \boldsymbol{x}_{-i})\} = \{\mathcal{S}_1(x_1, \boldsymbol{x}_{-1}), \ldots, \mathcal{S}_n(x_n, \boldsymbol{x}_{-n})\}$ is the set of surplus functions that each player $i \in \mathcal{N}$ wishes to maximize. Furthermore,

$$\mathcal{S}_i(x_i, \boldsymbol{x}_{-i}) = \mathcal{U}_i(x_i) - x_i \, p\left(\frac{\sum_{i=1}^{n} x_i}{n}\right). \quad (7)$$

Without loss of generality, for each user $i \in \mathcal{N}$, $\mathcal{S}_i(x_i, \boldsymbol{x}_{-i}) \geq 0, \forall x_i \geq 0$. Then we can infer that $r_i \geq \mu(x) = \frac{aq}{n}$. In game $G_1$, each user $i \in \mathcal{N}$ strategically selects its rate $x_i \geq 0$ to maximize its payoff function $\mathcal{S}_i(x_i, \boldsymbol{x}_{-i})$. From this model we can conclude that there may exist such stable strategies, which are identified as Nash Equilibria. A system is said to be in a Nash Equilibrium when no individual user $i \in \mathcal{N}$ can increase its own payoff by means of changing its strategy $x_i$ ultimately. A Nash equilibrium of game $G_1$ can be defined as a rate vector $\boldsymbol{x}^* \geq 0$, such that for all users $i \in \mathcal{N}$, $\forall i \geq 0$, we have $\mathcal{S}_i(x_i^*, \boldsymbol{x}_{-i}^*) \geq \mathcal{S}_i(x_i, \boldsymbol{x}_{-i}^*)$.

*B. Nash Equilibrium and POA*

We first show that a Nash equilibrium exists for this game under the new average congestion pricing mechanism. The interested readers can find the proof with reference to [8]. Therefore, we achieve that

(i) Game $G_1$ always has a unique Nash equilibrium.

(ii) Given $x^s$ is an optimal solution for function (3) and $x^*$ is a feasible Nash Equilibrium for game $G_1$, we want to get the character

$$\text{minimize POA} = \frac{\sum_{i=1}^n \mathcal{U}_i(x_i^*) - \sum_{i=1}^n \mathcal{P}(x_i^*)}{\sum_{i=1}^n \mathcal{U}_i(x_i^s) - \sum_{i=1}^n \mathcal{P}(x_i^s)}, \quad (8)$$

where $x^s$ and $x^*$ should satisfy $1 \geq x^s \geq 0$ and $1 \geq x^* \geq 0$, respectively. From the given conditions $\frac{\partial \mathcal{U}_i(x_i)}{\partial x_i} = r_i$ and $q = \sum_{i=1}^n x_i \leq 1$, the following inequality holds:

$$\frac{\sum_{i=1}^n \mathcal{U}_i(x_i^*) - \sum_{i=1}^n \mathcal{P}_i(x_i^*)}{\sum_{i=1}^n \mathcal{U}_i(x_i^s) - \sum_{i=1}^n \mathcal{P}_i(x_i^s)} \geq \frac{\min_{1 \geq x_i^* \geq 0}[\sum_{i=1}^n r_i x_i^* - \sum_{i=1}^n \mathcal{P}_i(x_i^*)]}{\max_{1 \geq q \geq 0}[(\max_i r_i)q - \sum_{i=1}^n \mathcal{P}_i(x_i^s)]}. \quad (9)$$

*Proposition 2*: Suppose that Assumptions 1, 2 and 3 hold, and that $p(z) = az$, for some $a > 0$. Also, $\mathcal{U}_i(0) \geq 0$ and $\max_i r_i = d$ are assumed. We prove that POA $\geq \frac{1}{2}$.

*Proof*: First, we show without loss of generality that $\mathcal{S}_i(x_i, \boldsymbol{x}_{-i}) \geq 0$, $\forall x_i \geq 0$ obtained before, so it is obvious that $\sum_i \mathcal{U}_i(x_i) \geq \sum_i \mathcal{P}_i(x_i)$. Subsequently, referring to [8], as known $1 \geq x_1 \geq x_2 \geq \ldots x_n \geq 0$, the optimal solution happens when the total rates (no more than fixed capacity 1) with all the $n$ users existing are only allocated to user 1: $x_1 = 1$, $x_2 = 0$, $x_3 = 0$, ..., $x_n = 0$, and $r_1 = d$. Consequently, the maximal aggregate surplus is $d - \frac{a}{n}$, when $n \to \infty$, $\sum_{i=1}^n \mathcal{U}_i(x_i^s) - \sum_{i=1}^n \mathcal{P}_i(x_i^s) = d$, thus the optimal solution of function (3) is obtained.

Next, the worst case occurs when the utility functions of the users are linear (Assumption 3). We then optimize over all games with linear utility functions to determine the worst case efficiency loss. The worst case of game $G_1$ is to solve the optimization problem as follows:

$$\text{minimize } \sum_{i=1}^n \mathcal{U}_i(x_i^*) - \sum_{i=1}^n \mathcal{P}_i(x_i^*), \quad (10)$$

where $x_i^*$ is a Nash equilibrium for Game $G_1$. Because the surplus $\mathcal{S}_i$ is concave for fixed $\boldsymbol{x}_{-i}$, a vector $x_i^*$ is a Nash equilibrium if and only if the following first-order conditions are satisfied for each $i$,

$$\frac{\partial \mathcal{S}_i(x_i, \boldsymbol{x}_{-i})}{\partial x_i} = \frac{\partial \mathcal{U}_i(x_i)}{\partial x_i} - \frac{\partial \mathcal{P}_i(x_i)}{\partial x_i} = r_i - \mu(x) - x_i \frac{\partial \mu(x)}{\partial x_i} = 0, (11)$$

$$r_i = \mu(x) + x_i \frac{\partial \mu(x)}{\partial x_i} = \frac{aq}{n} + \frac{ax_i}{n}, \; x_i > 0, \quad (12)$$

$$r_0 \leq \mu(x) = \frac{aq}{n}, \; x_i = 0. \quad (13)$$

Given $r_i \geq \mu(x) = \frac{aq}{n}$ before, then $r_0 = 0$ is concluded. To find the solution to (10), none of the user can be allocated the total capacity. We use these conditions to investigate the efficiency loss when users are price anticipating. Specifically, we are interested in comparing the aggregate surplus achieved at a Nash equilibrium with the aggregate surplus achieved at a optimal solution as (8). As a result, from (12) and the given condition $1 \geq x_1 \geq x_2 \geq \ldots x_n \geq 0$, it indicates that $d = r_1 \geq r_2 \ldots \geq r_n > 0$. Therefore, according to the Chebyshev's sum inequality, we have

$$\sum_{i=1}^n r_i x_i^* \geq \frac{1}{n} (\sum_{i=1}^n r_i)(\sum_{i=1}^n x_i^*), \quad (14)$$

where $\sum_{i=1}^n r_i = aq + \frac{aq}{n}$ owing to (12), and it is dependent on the total rates through the single link. Therefore, we obtain the aggregate surplus under the worst case of Nash Equilibria for game $G_1$: $\sum_{i=1}^n \mathcal{U}_i(x_i^*) - \sum_{i=1}^n \mathcal{P}(x_i^*) \geq \frac{aq^2}{n^2}$.

Hence, the POA, i.e., the efficiency boundary is

$$\text{POA} = \frac{\sum_{i=1}^n \mathcal{U}_i(x_i^*) - \sum_{i=1}^n \mathcal{P}(x_i^*)}{\sum_{i=1}^n \mathcal{U}_i(x_i^s) - \sum_{i=1}^n \mathcal{P}(x_i^s)} \geq \frac{\frac{aq^2}{n^2}}{d - \frac{a}{n}}. \quad (15)$$

For $\frac{2q^2+n}{n^2} \geq \frac{d}{a} \geq \frac{q^2+n}{n^2}$, when $n \to \infty$, the POA approaches 1/2, as required. The worst case efficiency loss is always exactly 50% for any ACS price function by an appropriate choice of utility and price functions. Furthermore, the corresponding numerical results on POA for 100 randomly sharing users with $q$ approaching to 1 are shown in Fig. 3. ∎

## III. ACS PRICING MECHANISM ---- NETWORK CODING AND ROUTING GAME

We also concentrate on the impacts on the selfish users with the ACS pricing mechanism when inter-session network coding is applied. Without loss of simplicity and generality, we only have considered two users performing the inter-session network coding as shown in Fig. 2.

The network model is similar to that in Fig. 1, except two aspects: 1) the two direct side links: $(u_1, v_n)$ from source node $u_1$ to destination node $v_n$, and $(u_n, v_1)$ from source node $u_n$ to destination node $v_1$; 2) shared single link divided into two independent single links from the view of two data transmission techniques (network coding and routing). In this scenario, the first and the last users (i.e., users 1 and $n$) can perform inter-session network coding. Let $D_1$ and $D_n$ denote the packets transmitted from sender nodes $u_1$ and $u_n$, respectively. The intermediate node $w$ can encode packets $D_1$ and $D_n$ together, and then send the encoded packet, denoted by $D_1 \oplus D_n$, towards node $z$ (and from there towards $v_1$ and $v_n$) through the network coding link. Given the remedy data $D_1$ from the side link $(u_1, v_n)$ and the remedy data $D_n$ from the side link $(u_n, v_1)$, nodes $v_n$ and

$v_1$ can decode the encoded packets that they receive. In fact, nodes $v_n$ and $v_1$ can both decode $D_1$ and $D_n$. What's more, the packet $D_i$ without encoded is simply transmitted from node $w$ to node $z$ through the routing link. Here $x_1$ and $x_n$ denote the data rates of source nodes $u_1$ and $u_n$, respectively. We have $x_1 \ne x_n$ by independent users 1 and $n$ in general.

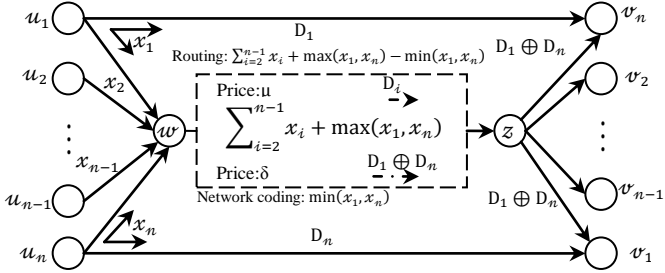

Figure 2. A single link shared by two independent flows, i.e., network coding and routing. Users 1 and $n$ perform inter-session network coding.

*Assumption 4*: The side links $(u_1, v_n)$ and $(u_n, v_1)$ in Fig. 2 always have zero cost and impose zero prices.

### A. Discriminatory prices

In contrast to the routing-only situation in section II, we establish two discriminatory prices $\mu$ and $\delta$ for network coding and routing users for each independent single link, respectively, as displayed in Fig. 2. Both the two prices are based on the ACS pricing mechanism in proposition 1.

Under such congestion network setting in Fig. 2, the routing group is combined by $n-2$ (assume there exist $n \ge 2$ users) routing-only users and the one between 1 and $n$ with rate $\max(x_1, x_n)$ (i.e., user 1, for the assumption $1 \ge x_1 \ge x_2 \ge \ldots x_n \ge 0$ ). In addition, they transmit data through the shared routing link at rate $x_i$ for each user $i \in \mathcal{N} \setminus \{1, n\}$ and at $[\max(x_1, x_n) - \min(x_1, x_n)]$ for user 1, respectively. As a result, the price for routing users following the ACS pricing mechanism is

$$\mu(x) = p\left(\frac{\sum_{i=2}^{n-1} x_i + \max(x_1, x_n) - \min(x_1, x_n)}{n-1}\right), \quad (16)$$

where $\sum_{i=2}^{n-1} x_i + \max(x_1, x_n) - \min(x_1, x_n)$ is the sum of rates through the routing link as shown in Fig. 2, and the variable $n-1$ is the number of users forwarding data on the routing link, i.e., $(n-2+1)$.

Afterwards, the price for network coded users following the same ACS pricing mechanism is defined as

$$\delta(x) = p\left(\frac{\min(x_1, x_n)}{2}\right) = p\left(\frac{x_n}{2}\right), \quad (17)$$

where $\min(x_1, x_n)$ is the sum of rates through the network coding link, the constant 2 is the number of users transmitting data on the network coding link, i.e., users 1 and $n$.

In particular, the aggregate data rate is no more than the fixed capacity 1 over the routing link of the single link $(w, z)$, i.e., $\sum_{i=2}^{n-1} x_n + \max(x_1, x_n) = q \le 1$ as a whole. Clearly, the benefit of the network coding is to reduce the traffic load on the shared link $(w, z)$ (thus reducing the link cost) and save the resources while achieving the same rates.

Moreover, we want to investigate the interactions between the two discriminatory prices depending on the changes of rates and the total number of users. Then we define

$$\beta = \frac{\delta(x)}{\mu(x)} = \frac{(n-1)x_n}{2(q - x_n)}. \quad (18)$$

From the equation, we can see that $\beta$ is only affected by the total number $n$ of users, the total rates $q$ through the whole link and the minimal network coding rate $x_n$. Owing to the assumption that $1 \ge x_1 \ge x_2 \ge \ldots x_{n-1} \ge x_n \ge 0$, we can conclude that $0 \le \beta \le \frac{1}{2}$ when $n \to \infty$, and $0 \le \beta < \infty$ when $n = 2$. Therefore, compared with [5] there exist wider ranges for $\beta$ because of the number of sharing users and their choices between network coding and routing.

### B. Game and POA

There are enough conditions to define the resource allocation game $G_2$ among network coding and routing flows as follows:

$$G_2 = \langle \mathcal{N}, \{x_i\}, \{S_i(x_i, \boldsymbol{x}_{-i})\}\rangle, \quad (19)$$

where $\mathcal{N} = \{1, 2 \ldots, n\}$ is the set of players (sharing users), $x_i$ is the strategy of each player $i \in \mathcal{N}$, and $\{S_i(x_i, \boldsymbol{x}_{-i})\} = \{S_1(x_1, \boldsymbol{x}_{-1}), S_2(x_2, \boldsymbol{x}_{-2}), \ldots, S_n(x_n, \boldsymbol{x}_{-n})\}$ is the set of surplus functions that each player $i \in \mathcal{N}$ wishes to maximize.

The network coding users 1 and $n$ have

$$S_1(x_1, \boldsymbol{x}_{-1}) = U_1(x_1) - (x_1 - x_n)\mu(x) - x_n\delta(x), \quad (20)$$

$$S_n(x_n, \boldsymbol{x}_{-n}) = U_n(x_n) - x_n\delta(x). \quad (21)$$

While each routing user $i \in \mathcal{N} \setminus \{1, n\}$ has

$$S_i(x_i, \boldsymbol{x}_{-i}) = U_i(x_i) - x_i\mu(x). \quad (22)$$

We also analyze the performance of two discriminatory prices by efficient allocation, and maximize the aggregate surplus as an optimal solution similar in (3). According to the similar proof in section II, the optimal solution of game $G_2$ occurs when the variable $\max_i(r_i)$ is allocated to all the users. That means

$$\sum_{i=1}^n S_i(x_i) \le dq + dx_n - \frac{a(q-x_n)^2}{n-1} - ax_n^2. \quad (23)$$

If $\frac{d}{a} \ge 2q$, the maximal value of $\sum_{i=1}^n S_i(x_i)$ is $2d - a$. It implies the optimal rate allocation that $x_1 = x_n = q = 1$ and $\sum_{i=2}^{n-1} x_i = 0$, which means the shared link only occupied by network coding users transmitting their data.

If $\frac{d}{a} < 2q \le 1$, the maximal value of $\sum_{i=1}^n S_i(x_i)$ is $\frac{4nad + (n-3)d^2}{4(n-1)a}$ when $q \ge x_n \ge \frac{d}{2a}$. By (11) and (12) in II, the worst case is identified by finding the worst Nash Equilibrium according to the equations (20), (21) and (22) in game $G_2$. Furthermore, the utility parameters for a Nash Equilibrium are $r_1 = \frac{a(q+x_1-2x_n)}{n-1}$, $r_n = ax_n$ for network coding user 1 and user $n$ respectively, and $r_i = \frac{ax_i}{n-1} + \frac{a(q-x_n)}{n-1}$ for each routing user $i \in \mathcal{N} \setminus \{1, n\}$.

Due to the above equations of $r_i$ for each user $i \in \mathcal{N}$, we can infer that $\sum_{i=1}^{n-1} r_i = \frac{an(q-x_n)}{n-1}$. The first-order derivative equation $\overline{\mathcal{U}_i(x_i)} = r_n$ is non-decreasing because of the concave function $\mathcal{U}_i(x_i)$. To obtain the worst case of game $G_2$, the same objective problem as in (10) is required to be solved. In this paper, only two exceptional cases at $n = 2$ and $n \to \infty$ are analyzed in detail.

First, we consider the situation at $n = 2$, only user 1 and user 2, for $1 \geq q = x_1 \geq x_2 \geq 0$, the aggregate surplus is: $\sum_{i=1}^{2} S_i(x_i) = dq - a(q - x_2)^2$. We obtain $x_2 \leq \frac{2q}{3}$, and $\frac{d}{a} \geq q$ owing to the given assumptions. Finally, the worst case is $dq - aq^2$ when $x_2 = 0$.

Then we consider the other situation at $n \to \infty$. We also suppose $1 \geq x_1 \geq x_2 \geq \ldots x_{n-1} \geq x_n \geq 0$, and then $d = r_1 \geq r_2 \geq \cdots \geq r_{n-1} \geq r_n > 0$ for the concave utility function, so we can get the following range $\frac{q}{n-1} \leq \frac{d}{a} \leq \frac{2(n-2)q}{(n-1)^2}$. With Chebyshev's sum inequality in (14), we can solve the aggregate surplus: $\sum_{i=1}^{n} S_i(x_i) \geq \frac{anq(q-x_n)}{(n-1)^2} - \frac{a(q-x_n)^2}{n-1}$.

Thus we can see that the worst case is a quadric equation of action $x_n$. For simplicity, let $x = q - x_n, \frac{(n-2)q}{n-1} \leq x \leq q$, we denote $g(x) = \frac{anqx}{(n-1)^2} - \frac{ax^2}{n-1}$. When the number of sharing users $n \to \infty$, $g(x)$ is non-decreasing, it indicates that $g(q) \leq g(x) \leq g\left(\frac{(n-2)q}{n-1}\right)$. Then we can obtain the worst case of game $G_2$ is the aggregate surplus $\frac{aq^2}{(n-1)^2} \leq \frac{d^2}{a}$ when $x = q$.

As mentioned above, we summarize the POA situations of different numbers of users sharing the single link when both network coding and routing are applied. The proof in section II can be as references.

When $n = 2$, $\frac{d}{a} \geq 2q \geq 2$, then

$$\text{POA} \geq \frac{1}{3}. \qquad (24)$$

When $n \to \infty$, $\frac{q}{n-1} \leq \frac{d}{a} \leq \frac{q}{2}$, then

$$\text{POA} \geq \frac{4}{9}. \qquad (25)$$

The above results extend the POA for routing-only flows in section II. In addition, the POA boundary is decreasing from $\frac{1}{3}$ to $\frac{4}{9}$ as the sharing users increasing from 2 to a quite large number with proper choices of price and utility parameters, $a$ and $d$, respectively. Compared with the routing-only boundary $\frac{1}{2}$ in section II, it suggests that network coding cannot improve the POA remarkably with the ACS.

## IV. CONCLUSION

This paper considers the ACS pricing model dealing with network resource allocation problem between two independent techniques for data transmission, i.e., routing and network coding. Without loss of generality, we focus on the case where two out of $n \geq 2$ users performing inter-session network coding in a sharing single link. The results are dramatically different from the case when the AMC is applied. The ACS pricing mechanism can be utilized to represent both the non-discriminatory pricing with routing-only users and discriminatory pricing with both routing and network coding simultaneously. We show that the choice of different transmission techniques and the number of sharing users can affect on the exact value of POA, but actually network coding cannot improve the POA obviously. Therefore, for more efficient use of limited resources, it indicates the sharing users have a higher tendency to choose network coding.

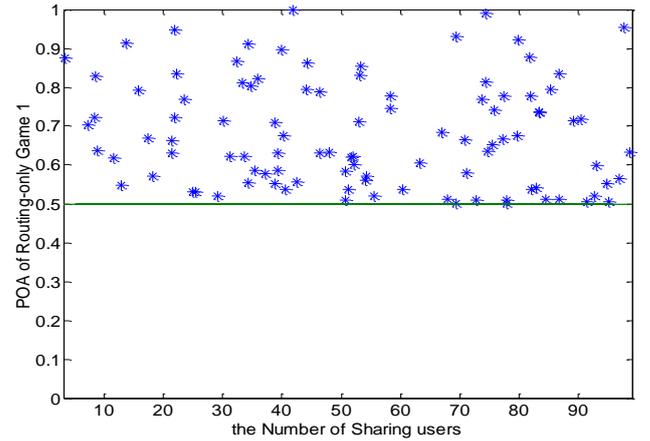

Figure 3. POA of 100 randomly routing-only sharing users.

REFERENCES

[1] R. Ahlswede, N. Cai, S. Li, and R. Yeung, "Network information flow,"IEEE Trans. on Information Theory, vol. 46, pp. 1204–1216, Apr. 2000.

[2] X. Liang, "Matrix games in the multicast networks: maximum information flows with network switching," IEEE Trans. on Information Theory, vol. 52, pp. 2433–2466, June 2006.

[3] Z. Li, "Cross-monotonic multicast," in Proc. of IEEE INFOCOM, Phoenix, AZ, Apr. 2008.

[4] X. Zhang and B. Li, "Dice: a game theoretic framework for wireless multipath network coding," in Proc. of ACM MobiHoc, Hong Kong, China, May 2008.

[5] A. Hamed Mohsenian-Rad, Jianwei Huang, and Vincent W.S. Wong, "A Game-Theoretic Analysis of Inter-Session Network Coding", IEEE ICC, Dresden, Germany, June, 2009

[6] Amir-Hamed Mohsenian-Rad, Jianwei Huang, Vincent W.S. Wong, and Robert Schober, "Repeated Inter-Session Network Coding Games: Efficiency and Min-Max Bargaining Solution", unpublished.

[7] Amir-Hamed Mohsenian-Rad, Jianwei Huang, Vincent W.S. Wong, "Inter-Session Network Coding with Strategic Users: A Game-Theoretic Analysis of Network Coding", unpublished.

[8] Ramesh Johari, and John N. Tsitsiklis, "A Scalable Network Resource Allocation Mechanism With Bounded Efficiency Loss", IEEE Journal On Selected Areas In Communications, Vol. 24, NO. 5, May 2006.

[9] Vasilis Ntranos, "Cost Sharing Network Routing Game", www-bcf.usc.edu/~shanghua/teaching/Fall2010/cost_sharing.pdf.